\newcommand{\et}{$\eta$~Carinae}
\documentclass[]{aa}
\usepackage{natbib,graphics,graphicx,psfig}
\bibpunct{(}{)}{;}{a}{}{,}

\begin{document}

\headnote{Letter to the Editor}

\title{Direct measurement of the size and shape of the present-day
stellar wind of \et
\thanks{Based on observations obtained at the European Southern Observatory}}
\author{R. van Boekel \inst{1,2} \and P. Kervella \inst{1} \and M.
Sch\"oller\inst{1} \and T. Herbst \inst{3} \and W. Brandner \inst{1,3}
\and A.~de~Koter \inst{2} \and L.B.F.M.~Waters \inst{2,4} \and D.J.
Hillier \inst{5} \and F. Paresce \inst{1} \and R. Lenzen \inst{3} \and
A.-M. Lagrange \inst{6}}

\institute{
European Southern Observatory,
Karl-Schwarzschild-Strasse 2, D-85748 Garching, Germany
\and
Astronomical Institute ``Anton Pannekoek'', University of
Amsterdam, Kruislaan 403, 1098 SJ Amsterdam, The Netherlands
\and
Max-Planck-Institut f\"ur Astronomie, K\"onigstuhl
17, 69117 Heidelberg, Germany
\and
Instituut voor Sterrenkunde, K.U.~Leuven, Celestijnenlaan
200B, 3001 Heverlee, Belgium
\and
Department of Physics and Astronomy, University of Pittsburgh,
3941 O'Hara Street, Pitssburgh, PA 15260, U.S.A.
\and
Laboratoire d'Astrophysique de l'Observatoire de Grenoble,
Universit\'e J. Fourier, CNRS, BP 53, 38041 Grenoble Cedex 9, France}

\offprints{R. van Boekel:\\ vboekel@science.uva.nl}

\date{Received $<$date$>$; accepted $<$date$>$}

\authorrunning{Van Boekel et al.}
\titlerunning{Direct measurement of size and shape of \et \ wind}

\abstract{We present new high angular resolution observations at
near-IR wavelengths of the core of the Luminous Blue Variable \et,
using NAOS-CONICA at the VLT and VINCI at the VLT Interferometer
(VLTI). The latter observations provide spatial information on a scale
of 5 milli-arcsec or $\sim$11 AU at the distance of \et. The
present-day stellar wind of \et \ is resolved on a scale of several
stellar radii.  Assuming spherical symmetry, we find a mass loss rate
of 1.6$\times$10$^{-3}$ M$_{\odot}$/yr and a wind clumping factor of
0.26. The VLTI data taken at a baseline of 24 meter show that the
object is elongated with a de-projected axis ratio of approximately
1.5; the major axis is aligned with that of the large bi-polar nebula
that was ejected in the 19th century. The most likely explanation for
this observation is a counter-intuitive model in which stellar
rotation near the critical velocity causes enhanced mass loss along
the rotation axis.  This results from the large temperature difference
between pole and equator in rapidly rotating stars. \et \ must rotate
in excess of 90 per cent of its critical velocity to account for the
observed shape.  The large outburst may have been shaped in a similar
way. Our observations provide strong support for the existence of a
theoretically predicted rotational instability, known as the $\Omega$
limit.}

\maketitle


\section{Introduction}

The Luminous Blue Variable \et \ is the most luminous star known in the
galaxy \citep{1997ARA&A..35....1D}. Its extreme properties make it an
interesting laboratory to study the physics of the most massive stars
in galaxies. Not much is known about the life of such stars, including
their birth and post-main-sequence evolution. \et \ has already left
the main sequence and is now in an unstable phase; it experienced a
large outburst in the 19th century which created the beautiful bi-polar
nebula seen in many images (referred to as the homunculus). As much as
10 M$_{\odot}$ may have been ejected during that event
\citep{2003AJ....125.1458S}. A smaller outburst seems to have occurred
around 1890, creating a smaller, but similarly shaped nebula hidden
inside the larger one \citep{2003AJ....125.3222I}.

The origin of the highly bi-polar shape of the homunculus is a matter
of debate. Model calculations show that a spherical explosion into a
massive equatorial torus can explain the observed geometry
\citep{1995ApJ...441L..77F}. Indeed, evidence for the presence of a 15
M$_{\odot}$ torus was found from ISO spectroscopy
\citep{1999Natur.402..502M}. Note however, that \cite{2003AJ....125.1458S}
argue that most of this mass is actually located in the lobes. Other
models reproduce the shape of the nebula by assuming a non-spherical
outburst that runs into a spherical envelope. Recently, several authors
proposed that luminous stars rotating close to their critical speed
have stellar winds with a higher wind density and expansion velocity
\emph{at the poles} (Owocki et al.~\citeyear{1996ApJ...472L.115O}, 
Maeder \& Desjacques~\citeyear{2001A&A...372L...9M},
Dwarkadas \& Owocki~\citeyear{2002ApJ...581.1337D}).
Therefore, the shape of the homunculus may
be a natural consequence of the combined effects of rapid rotation and
the high luminosity of \et. Note that the extreme luminosity of \et \
implies that even a modest amount of rotation causes the star to be
close to its critical velocity \citep{1997lbv..conf...83L}. Evidence in
support of a polar enhanced wind was recently inferred from Hubble
Space Telescope (HST) spectroscopy of starlight scattered off dust
grains in the homunculus (Smith et al.~\citeyear{2003ApJ...586..432S}).

In this \emph{Letter} we present the first results of an extensive high
angular resolution near-IR study of the core of the homunculus,
revealing for the first time the shape of the present-day stellar wind
on a scale of 5 milli-arcsec.

\section{Observations}

\begin{figure}
\resizebox{\hsize}{!}{\includegraphics{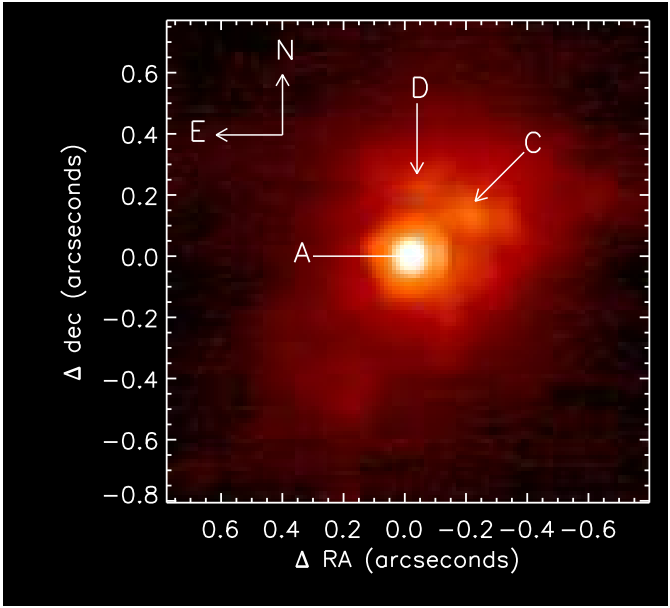}}
\caption[]{Diffraction-limited 2.29 $\mu$m image of the core of the \et \
nebula, taken with the adaptive optics camera NAOS-CONICA at the VLT.
The position of the Weigelt Blobs is indicated.} \label{fig:naco}
\end{figure}

We have observed \et \ with the adaptive optics camera NAOS-CONICA
(Lenzen et al.~\citeyear{1998SPIE.3354..606L}, 
Rousset et al.~\citeyear{2000SPIE.4007...72R}) attached to
Yepun, one of the 8.2 meter Unit Telescopes of the Very Large Telescope
(VLT) of the European Southern Observatory (ESO), located at Cerro
Paranal, Chile. NAOS was in the visual wavefront sensor configuration
with 14 $\times$ 14 subapertures used for wavefront sensing. CONICA was
used with the S13 camera (13.25\,milli-arcsec/pixel) and the NB\_239
intermediate band filter with a central wavelength of 2.39\,$\mu$m and
a width of 60\,nm. We employed a neutral density filter with an
attenuation factor of $\approx$70 at the observing wavelength in order
to avoid saturating the central peak of the point spread function.
Individual exposures were 0.16\,s long, and 20 exposures were co-added,
resulting in a total exposure time of 3.6\,s per frame. Four such
frames were obtained. The reduced image, diffraction limited at 70
milli-arcsec, is shown in figure \ref{fig:naco}.

The central 1.5 arcsecond region of the nebula is dominated by a point
source and shows a complex morphology of blobs in the immediate
vicinity. The brightest blobs were previously discovered using speckle
imaging techniques (Weigelt \& Ebersberger~\citeyear{1986A&A...163L...5W},
Hofmann \& Weigelt~\citeyear{1988A&A...203L..21H}). 
There is a clear asymmetry in the
emission, which is brightest in the north-west. Approximately 57 per
cent of the flux seen inside a region with a diameter of 1.4 arcsec,
centered on the core of the nebula and corresponding to the Airy disk
of the siderostats, is concentrated in the unresolved point source.  We
use recent literature data \citep{2000ApJ...529L..99S} to flux
calibrate our observations and find that the unresolved central source
has a flux of about 200~Jy.

We used the two 35~cm test telescopes of the VLT Interferometer 
(Glindemann et al.~\citeyear{2003SPIE.4838...89G})
and the test instrument VINCI \citep{2000SPIE.4006...31K} to obtain
interferometric measurements at baselines ranging from 8 to 62 meters
in length. The observations were carried out in the first half of 2002
in four different nights, and again in early 2003. For a full
description of the observations, and analysis of the data we refer to
Sch\"oller et al. (submitted to A\&A). The baselines used, have a ground 
length of 8, 16, 24, and 66\,m respectively. In particular observations 
with the 24\,m baseline cover a wide range of projected 
baseline orientations.

In figure~\ref{fig:vinci} we show the VINCI observations, given as
squared visibility points. The visibility curves were corrected for
the emission in the siderostat beams on scales larger than 70
milli-arcsec, by adding the resolved component seen in our NAOS-CONICA
images.  This results in a reduction of the visibilities by a factor
(1-f$_{\mathrm{res}}$), where f$_{\mathrm{res}}$ is the fraction of
the total flux in the siderostat beams that does not come from the
central source (0.43).  For each baseline, the top panel of
figure~\ref{fig:vinci} shows the average visibilities at that
baseline, while the bottom panel shows the variations of visibility
with projected baseline orientation for the 24m baseline.

\section{Analysis}

\subsection{Nature of the emission}

Figure~\ref{fig:vinci} shows that the visibility decreases with
increasing baseline, indicating that the interferometer is resolving
the central source. The near-IR emission seen in the central region of
\et \ must come from an object that emits half of its flux in an area
of 5 milli-arcsec or 11 AU diameter, assuming circular
symmetry and adopting a distance of 2300 parsec
\citep{1997ARA&A..35....1D}. Using the measured size and flux of 200~Jy
\citep{2000ApJ...529L..99S}, a lower limit to the temperature of the
object can be derived, assuming that at a wavelength of 2.2 $\mu$m
there is no extinction. We find a minimum temperature of 2300~K.
Allowing for a foreground extinction of one (two) magnitudes
\citep{2001ApJ...553..837H}, this temperature goes up to 3200 (5000)~K.
The minimum temperature is too high to be caused by thermal emission from
dust. We conclude that we have spatially resolved the ionized stellar
wind.

\subsection{Mass loss rate}

We constructed a simple physical model for the observed size and flux
of \et, assuming that it is due to a spherical star with a dense,
ionized and isothermal stellar wind that reaches a terminal flow
velocity of 500 km/s \citep{2001ApJ...553..837H}.  The wind is further
assumed to be clumpy, with the clumping factor $f$ defined as
$\overline{\rho} = f~\rho$ and where $\overline{\rho}$ is the unclumped
wind density. For a more detailed description of the model assumptions,
we refer to de Koter et al. (in preparation).

\begin{figure}
\resizebox{\hsize}{!}{\includegraphics{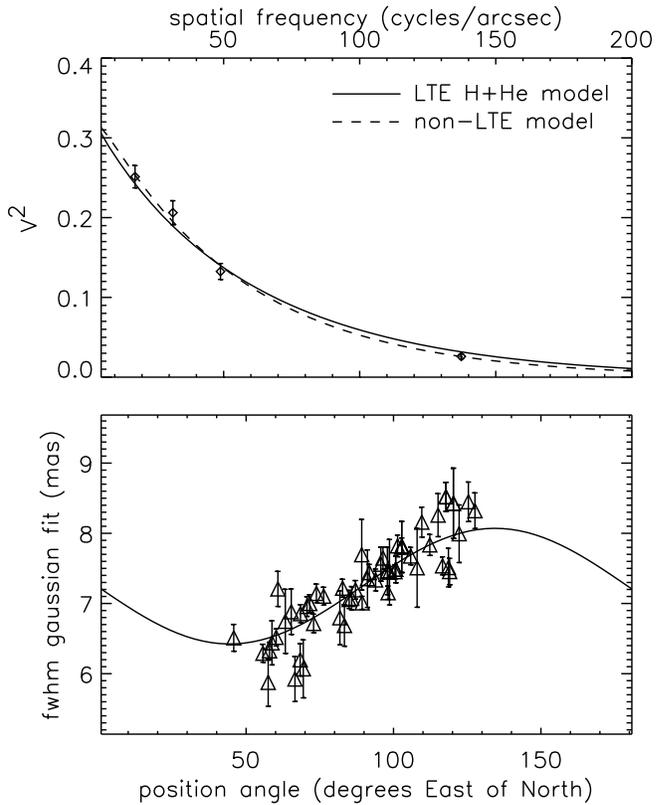}}
\caption[]{\emph{upper panel}: Observed visibility as a function of
projected baseline. The full line is the best fit LTE wind model to
the VLTI observations. The dashed line is the best fit non-LTE
radiative transfer model to the observed visibilities. \emph{Lower
panel:} Variation of FWHM fitted to the visibility of \et \ measured
with VINCI at the VLTI as a function of projected orientation of the
24\,m baseline.  The solid line gives the best fit to the
measurements, assuming a 2D gaussian shape of the source at each
projected baseline orientation. The amplitude of the size variations
gives a ratio of major to minor axis of 1.25$\pm$0.05.  The major axis
has a position angle of 134$\pm$7 degrees east of north.}
\label{fig:vinci}
\end{figure}

\begin{figure}
\resizebox{\hsize}{!}{\includegraphics{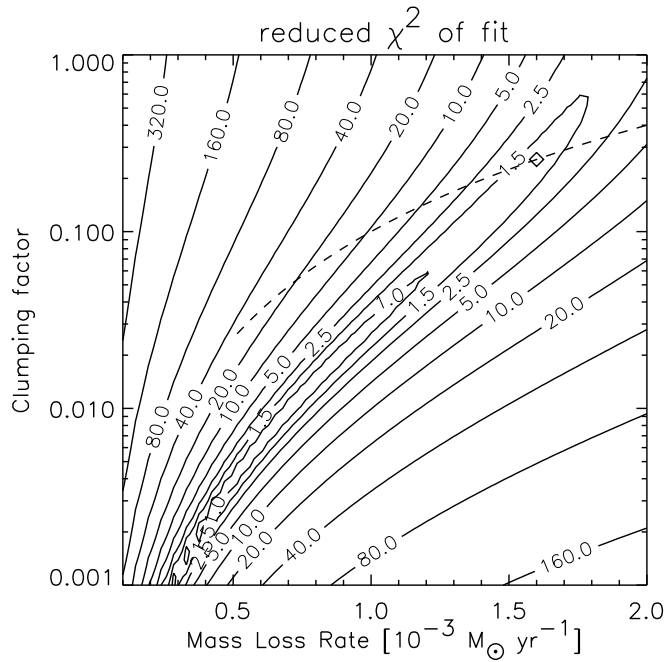}} \caption[]{
Allowed ranges in mass loss rate and wind clumping factor for the LTE
wind model (contours) and the full non-LTE radiative transfer model
applied to the Hubble Space Telescope spectra
\citep{2001ApJ...553..837H} (dashed line). Both data sets agree for a
mass loss rate of 1.6$\times$10$^{-3}$ M$_{\odot}$/yr and a wind clumping
factor of 0.26.} \label{fig:model}
\end{figure}

A range of mass loss rates and clumping factors are consistent with the
data. When combined with optical spectra of the central star taken with
the HST \citep{2001ApJ...553..837H}, a unique
determination of the mass loss rate and the wind clumping factor is
possible. The HST spectrum can be fitted using a non-local
thermodynamic equilibrium (non-LTE) model for the star and wind
\citep{2001ApJ...553..837H} which is degenerate in the ratio
$\dot{M}/\sqrt{f}$, where $f$ is the wind clumping factor as defined
above. A good fit to the HST observations is obtained for a ratio
$\dot{M}/\sqrt{f}$ = 0.00316 M$_{\odot}$/yr. We finally arrive at a
mass loss rate of 1.6$\pm$0.3$\times$10$^{-3}$ M$_{\odot}/$yr and a wind
clumping factor $f$~=~0.26 (figures~\ref{fig:vinci} and
\ref{fig:model}). The spatial light distribution of the non-LTE model
yields a somewhat better fit to the data than the LTE model.

\subsection{Asymmetry of the emission}

The 24\,m baseline (figure~\ref{fig:vinci}) data show a variation of
visibility with projected baseline orientation, which is inconsistent
with a spherical distribution of light at 2.2~$\mu$m. Deviations from
spherical symmetry will introduce additional uncertainty in the
derivation of the mass loss rate given above. To determine the axis
ratio of the emission, a gaussian shape was fit to the observations at
each projected baseline orientation. Note that the actual light
distribution is not gaussian. Therefore, the typical size of 7
milli-arcsec derived from the gaussian fits differs from the 5
milli-arcsec derived from our (spherical) physical model. We find a
major to minor axis ratio of 1.25 $\pm$ 0.05, and a position angle of
134 $\pm$ 7 degrees. This position angle is within errors equal to the
position angle of 132 degrees \citep{2001AJ....121.1569D} found for the
homunculus. We therefore have conclusive evidence that the density
contours in $\eta$~Carinae's wind are elongated along the major axis of
the Homunculus.  The chance that the two prolate structures - i.e. the
core and the Homunculus - have different orientation in 3 dimensions
but still show the observed alignment when projected on the plane of
the sky is about 10 percent, given the uncertainties in our
measurements. The core on a scale of 10$^1$ AU, and the
Homunculus reaching out several times 10$^4$ AU, must
be truly aligned in space. Adopting an angle between the major axis of
the Homunculus and the line of sight of 41 degrees
\citep{2001AJ....121.1569D}, the observed aspect ratio of major over
minor axis of 1.25 implies a de-projected aspect ratio of roughly 1.5
(in detail this de-projection depends on the actual three dimensional
structure of the object).

\section{Discussion}

It has been thought (e.g. Lamers \& Pauldrach
\citeyear{1991A&A...244L...5L}, Poe \& Friend~\citeyear{1986ApJ...311..317P})
that stellar rotation enhances mass loss in
the equatorial regions, resulting in disk-like winds. However, this
would imply that $\eta$ Carinae's rotation axis is perpendicular to the
major axis of the bi-polar homunculus, which is unlikely. A recent
model (Owocki et al.~\citeyear{1996ApJ...472L.115O}, 
Dwarkadas \& Owocki~\citeyear{2002ApJ...581.1337D})
for line-driven winds from luminous
hot stars rotating near their critical speed predicts a higher wind
speed and density along the poles than in the equator. This
counter-intuitive effect is caused by the increased polar temperature
\citep{1924MNRAS..84..665V} and associated radiation pressure. Our data
clearly favour the polar wind model.  The VLT data do not provide
information about the velocity field. However, recent HST spectroscopy
\citep{2003ApJ...586..432S} of starlight reflected by dust in the
Homunculus indicates a latitude-dependent wind outflow velocity, with
the highest velocities near the pole; this is expected for a wind which
is stronger at the poles. These data also suggest a polar enhanced
wind density. Applying the model of Dwarkadas \& Owocki
(\citeyear{2002ApJ...581.1337D}, see also Maeder \& 
Desjacques~\citeyear{2001A&A...372L...9M}), we find
that \et \ must rotate at about 90 per cent of its critical velocity to
account for the observed shape.

The question arises whether the model assumptions (line-driven wind,
radiative envelope) made by e.g. \cite{2002ApJ...581.1337D} are
applicable to \et. We note that \et \ is almost certainly fully
convective \citep{1997lbv..conf...83L}, due to its near-Eddington
luminosity (defined as the luminosity where surface gravity is
compensated by radiation pressure). Therefore the difference between
polar and equatorial temperatures in \et \ will likely be smaller
\citep{1967ZA.....65...89L} than in polar wind models, that adopt
radiative envelopes. For such a convective
envelope to produce a substantial temperature contrast, \et \ must
rotate in excess of 0.9 of the critical speed. The observed mass loss
rate of 1.6$\times$10$^{-3}$ M$_{\odot}$/yr can be explained in terms of
radiation driven wind theory (C. Aerts, private communication).

The alignment of the homunculus and the present-day wind suggests a
common physical cause. Rotation may then also be responsible for the
shape of the homunculus \citep{2002ApJ...581.1337D}. An outburst in
1890 probably produced a bipolar nebula with a present-day size of 2
arcsec \citep{2003AJ....125.3222I}, which is aligned with, and inside
the larger homunculus. There is thus strong evidence that the wind
geometry is similar over a wide range of mass loss rates. It is not
likely however that line-driven wind models are applicable to the
outbursts. Nevertheless, our data underpin the importance of rotation for
the post-main-sequence evolution of very massive stars such as \et; it
seems inevitable that as the star evolves, it will run into a
rotational instability, referred to as the $\Omega$ limit
\citep{1999ApJ...520L..49L}.

\acknowledgements{We would like to thank C. Aerts, P. Morris and N.
Langer for useful discussions. DJH acknowledges partial support from 
NASA grant NAGW 3838.}

\bibliographystyle{aa}
\bibliography{references}

\end{document}